\documentclass[aps,prb,twocolumn,showpacs,superscriptaddress]{revtex4}
\usepackage{graphicx}
\usepackage{bm}

\begin{document}
\newcommand{\beq}{\begin{equation}}
\newcommand{\eeq}{\end{equation}}

\title{Classical
behavior of strongly correlated Fermi systems near
a quantum critical point. Transport properties}
\author{V.~A.~Khodel}
\affiliation{Russian Research Centre Kurchatov
Institute, Moscow, 123182, Russia}
\affiliation{McDonnell Center for the Space Sciences \&
Department of Physics, Washington University,
St.~Louis, MO 63130, USA}
\author{J.~W.~Clark}
\affiliation{McDonnell Center for the Space Sciences \& Department
of Physics, Washington University, St.~Louis, MO 63130, USA}
\author{M.~V.~Zverev}
\affiliation{Russian Research Centre Kurchatov
Institute, Moscow, 123182, Russia}
\affiliation{Moscow Institute of Physics and Technology, Moscow, 123098, Russia}
\date{\today}
\begin{abstract}
The low-temperature kinetics of the strongly correlated electron liquid
inhabiting a solid is analyzed.  It is demonstrated that a softly damped branch of transverse zero sound emerges when several bands cross the Fermi surface simultaneously near a quantum critical point
at which the density of states diverges.  Suppression of the
damping of this branch occurs due to a mechanism analogous to
that affecting the phonon mode in solids at room temperature,
giving rise to a classical regime of transport at extremely low
temperatures in the strongly correlated Fermi system.
\end{abstract}
 \maketitle

After a decade of comprehensive studies,\cite{loh,steglich} the
prevalence of non-Fermi-liquid (NFL) behavior in strongly correlated Fermi systems is no longer a revelation.  However, various features of NFL phenomena still await satisfactory explanation, especially
the puzzling observations pointing to characteristic classical
behavior in a quantum-critical regime.  For example, at extremely
low temperatures around 1 mK, experimental measurements of the
specific heat of two-dimensional (2D) $^3$He, as realized in dense
$^3$He films,\cite{saunders1,saunders2} are described by
the classical formula $C(T)=\beta+\gamma T$, where $\beta$
and $\gamma$ are constants.  Also in contrast to FL theory,
the low-temperature resistivity $\varrho(T)$ of many high-$T_c$
compounds in strong magnetic fields, as well as certain
heavy-fermion metals, is observed\cite{stegacta,stegcol,taillefer}
to vary linearly with $T$. Hence these systems behave as if
a major contribution to the collision term comes from the
electron-phonon interaction, in spite of the fact that the
phonon Debye temperature $T_D$ exceeds measurement temperatures
by a factor $10^2-10^3$.

Previously,\cite{kcsz} we have attributed the presence of the classical term $\beta$ in the specific heat $C(T)$ of 2D liquid $^3$He
to softening of the transverse zero-sound mode (TZSM), occurring
in the region of the quantum critical point\cite{saunders2} (QCP) where the density of states $N(0)$, proportional to the effective mass
$M^*$, diverges.  Here we shall address the impact of the TZSM on
transport properties in the QCP regime.  The TZSM exists only
in those correlated Fermi systems where the effective mass $M^*$
exceeds the bare mass $M$ by a factor more than 3, a requirement
that is always met on approaching the QCP.  In conventional
Fermi liquids, the Fermi surface consists of a single sheet,
so the TZSM has a single branch with velocity $c_t$ exceeding
the Fermi velocity $v_F$.  Consequently, there is a ban against
emission and absorption of sound quanta by electrons, and the
role of the TZSM in kinetics is of little interest.  However,
in heavy-fermion metals, it is usual for several bands to cross
the Fermi surface simultaneously, thereby generating several
zero-sound branches.  For all branches but one the sound velocities
are less than the largest Fermi velocity.   Hence the aforementioned
ban is lifted, and these branches of the TZSM spectrum experience
damping, in a situation similar to that for zero-spin sound.
In the latter instance, Landau damping is so strong that the mode
cannot propagate through the liquid.\cite{pines,halat}  It will
be seen, however, that this is not the case for damping of the
TZSM, because of the softening of this mode close to the QCP.
Due to the softening effect, the contribution of the damped
TZSM to the collision integral has the same form as the electron-phonon interaction at room temperature.  On the other hand, we will
also find that in heavy-fermion metals, analogously to the case of
liquid $^3$He films, softening of the TZSM acts to lower a characteristic temperature $\Omega_t$ that plays the role of a Debye temperature, setting the stage for the existence of a classical transport regime at extremely low temperatures.

In the canonical view of quantum phase transitions, the QCP
has been identified with an end point of the line $T_N(\rho)$
of a second-order phase transition associated with violation
of some Pomeranchuk stability condition.  In turn this violation
is associated with divergence of the energy derivative
$\partial\Sigma(p,\varepsilon)/\partial\varepsilon$ of the
self-energy and consequent vanishing of the quasiparticle weight
$z=(1-\partial\Sigma(p,\varepsilon)/\partial\varepsilon)^{-1}$
in single-particle states at the Fermi surface, thus
triggering\cite{loh,steglich,chubukov} divergence of the
effective mass $M^*$ defined by $M/M^* = z(1+\left(\partial
\Sigma({\bf p},\varepsilon) /\partial\epsilon^0_p \right)|_{p=p_F}$.

A number of key experimental studies performed recently fail to
support the canonical view of the QCP.  In 2D liquid $^3$He,
experiment\cite{saunders1,saunders2} has not identified any phase
transition that can be so associated with the point of the divergence
of  $N(0)$.  It has been acknowledged\cite{bud'ko,stegcol} that a similar situation (cf.~Ref.~\onlinecite{kczjetp09,ckzjmpb10}) also prevails for the QCPs of heavy-fermion metals.  In essence, the point where the density of states diverges is {\it separated by an intervening NFL phase} from points where lines of some second-order phase transition terminate. Furthermore, these transitions possess unusual properties such as hidden order parameters; therefore within the standard collective scenario they can hardly qualify as triggers of the observed rearrangements.

We are therefore compelled to interchange the horse and the cart, relative to the canonical scenario.  Following Refs.~\onlinecite{shagh,prb2005},
we attribute the QCP to vanishing of the Fermi velocity $v_F$ at
a critical density $\rho_{\infty}$, which occurs if $ 1+\left(\partial
\Sigma({\bf p},\varepsilon)/\partial \epsilon^0_p\right)|_{p=p_F}=0$.
Accordingly, in this scenario for the QCP, it is the
{\it momentum-dependent part} of the mass operator that plays
the decisive role.

The authors of most theory articles devoted to the physics of the
QCP claim that switching on the interactions between particles never
produces a significant momentum dependence in the effective interaction
function $f$, and hence that the option we propose and develop
is irrelevant.  This assertion cannot withstand scrutiny.  The
natural measure of the strength of momentum-dependent forces
in the medium is provided by the dimensionless first harmonic
$F_1=f_1p_FM^*/\pi^2$ of the interaction function
$f({\bf p}_1,{\bf p}_2)$ of Landau theory. In a system such
as 3D liquid $^3$He where the correlations are of moderate
strength, the result $F_1 \geq 6.25$ for this measure extracted
from specific-heat data is already rather large.  The data on
2D liquid $^3$He are yet more damaging to the claim of minimal
momentum dependence, since the effective mass is found to
{\it diverge} in dense films.\cite{godfrin1,saunders1,saunders2}
In the case of QCP phenomena occurring in strongly correlated
systems of {\it ionic} crystals, it should be borne in mind that the electron effective mass is greatly enhanced due to electron-phonon interactions that subserve polaron effects.\cite{pekar,alex,alexk}

The change in sign of $v_F$ at the QCP results not only in a
divergent density of states, but also in a {\it rearrangement of the
Landau state} beyond the QCP.  As a rule, however, such a
rearrangement already occurs {\it before} the system attains a QCP.
This may be understood from simple arguments based on the Taylor
expansion of the group velocity $v(p)=\partial\epsilon(p)/\partial p$,
which has the form
\beq
v(p)= v_F(\rho)+ v_1(\rho){p-p_F\over p_F} +{1\over 2}v_2{(p-p_F)^2
\over  p^2_F}
\label{vt}
\eeq
in the vicinity of the QCP.  We assert that the last coefficient $v_2$ is positive, to ensure that the spectrum
\beq
\epsilon(p)=v_F(\rho)(p-p_F)+ {1\over 2p_F}v_1(\rho)(p-p_F)^2
+{1\over 6p^2_F}v_2(p-p_F)^3
\label{spt}
\eeq
derived from Eq.~(\ref{vt}) exhibits proper behavior at
large separations from the Fermi surface.

By its definition, the QCP is situated at a density $\rho_\infty$
where $v_F(\rho)$ vanishes.  The QCP must in fact correspond to
an extremum of the function $v(p,\rho_\infty)$, which vanishes
for the first time at $p=p_F$.  Thus, the simultaneous vanishing
of the coefficient $v_1(\rho_\infty)= (dv(p,\rho_\infty)/dp)|_{p=p_F}$
of the second term in the Taylor series is {\it crucial} to the
occurrence of the QCP.  Generally, $v_1$ does not meet this additional
requirement. However, in relevant cases its finite value remains
extremely small, making it possible to tune the QCP by imposing
an external magnetic field.

When $v_1 \neq 0$ in Eq.~(\ref{spt}), this equation unavoidably
acquires two additional {\it real} roots at a critical density
$\rho_t$ where $v_F(\rho_t)=3v^2_1/(8v_2)$, namely
\beq
p_{1,2}-p_F= -p_F{3v_1\over 2v_2}\left(1\pm
\sqrt{1-{8v_F(\rho)v_2\over 3v^2_1}}\right).
\label{rootv}
\eeq
Clearly, this transition, identified as a topological phase
transition \cite{lifshitz,zb,shagp,volrev,prb2008}
(TPT), takes place {\it already}
on the disordered side of the QCP regime where $v_F(\rho)$ is still
positive.  Accordingly, a new hole pocket opens and the Fermi surface
gains two additional sheets, the new $T=0$ quasiparticle
momentum distribution being given by $n(p)=1$ for $p<p_1$ and
$p_2<p<p_F$, and zero elsewhere.  The emergence of new small
pockets of the Fermi surface is an integral feature of the QCP
phenomenon, irrespective of whether the strongly correlated Fermi
system is 2D liquid $^3$He, a high-$T_c$ superconductor, or a
heavy-fermion metal.

At the TPT point $\rho_t$, the density of states, given by
\beq
N(T)=\int {\partial n({\bf p},T)\over \partial \epsilon({\bf p})} d\upsilon
\label{ds}
\eeq
with $d\upsilon$ denoting an element of momentum space,
is also divergent.  Inserting the spectrum (\ref{spt}), straightforward calculation yields $N(T\to 0)\propto T^{-1/2}$, in contrast to
the behavior $N(T\to 0,\rho_\infty)\propto T^{-2/3}$ obtained\cite{prb2005} in the case where $v_1(\rho_\infty)=0$.

Having tracked the initial evolution of the topology of the Fermi surface in the QCP region, our analysis turns next to the salient features of the TZSM spectrum in systems having a multi-connected (i.e., multi-sheet) Fermi surface.  We first examine how the TZSM softens in 3D systems with a singly-connected Fermi surface, where the
dispersion relation has the well-known form\cite{halat}
\beq
1={F_1\over 6}\left[1- 3(s^2-1)\left({s\over 2}
\ln {s+1\over s-1}-1\right)\right]
\label{dr1}
\eeq
with $s=c/v_F$ and $F_1=f_1p_FM^*/\pi^2$.  The TZSM is seen to
propagate only if $F_1>6$, i.e., $M^*>3M$.  Near the QCP
where $M^*(\rho)\to\infty$, one has $v_F/c\to 0$, and
Eq.~(\ref{dr1}) simplifies to
\beq
1={F_1\over 15}{v^2_F\over c^2},
\label{dr11}
\eeq
which implies
\beq
c(\rho\to \rho_{\infty})\to \sqrt{{p_Fv_F(\rho)\over  M}}
\propto {p_F\over M}\sqrt{{M\over M^*(\rho)}}\to 0,
\label{ctr}
\eeq
an analogous formula being obtained for a 2D system.

To facilitate analysis of damping of the TZSM in systems having
a {\it multi-connected} Fermi surface, we restrict consideration
to the case of two electron bands.  The TPT
is assumed to occur at one of the bands, so that its Fermi velocity,
denoted again by $v_F$, tends to zero, while the Fermi velocity
$v_o$ of the other band remains unchanged through the critical
density region.  The model dispersion relation for the complex
sound velocity $c = c_R+ic_I$ becomes
$$
1={F_1\over 6}\left[ 1-3\left({c^2\over v^2_F}-1\right)
\left({c\over 2v_F}\ln {c+v_F\over c-v_F}-1\right)\right]+$$
\beq
+{F_1\over 6}{v_F\over v_o}\left[1-3\left({c^2\over v^2_o}-1\right)
\left({c\over 2v_o}\ln {c+v_o\over c-v_o}-1\right)\right].
\label{dr2}
\eeq
It can easily be verified that the contribution of the second term to
the real part of the right-hand side of Eq.~(\ref{dr2}) is small
compared to that of the first term, since $v_F/v_o\to 0$ toward the QCP. On the other hand, noting that
$\ln \left[(c_R+ic_I+v_o)/ (c_R+ic_I-v_o)\right] \simeq - i \pi$,
the corresponding contribution  $i\pi F_1v_Fc_R/(4v^2_o)$
to the imaginary part of the right-hand side cannot be ignored,
else $c_I=0$.  By this reasoning, Eq.~(\ref{dr2}) assumes the
simplified form
\beq
1={F_1\over 15}{v^2_F\over (c_R+ic_I)^2}- i{\pi\over 4v^2_o} F_1 v_Fc_R
\eeq
analogous to Eq.~(\ref{dr11}).  Its solution obeys
\beq
c_R\propto \sqrt{{M\over M^*(\rho)}}, \quad c_I\propto {M\over M^*(\rho)}.
\label{rvel}
\eeq
Importantly, we see then that the ratio $c_I/c_R\propto\sqrt{M/M^*(\rho)}$
is {\it suppressed} in the QCP regime, which allows us to analyze
the contribution of the TZSM to the collision term entering the resistivity along the same lines as in the familiar case of the electron-phonon interaction. By contrast, the group velocities of the damped branches of {\it longitudinal} zero sound are found to be
insensitive to variation of the effective mass in the
QCP region.\cite{pines,halat}  No such quenching by a small
parameter $M/M^*$ arises, so these modes cannot propagate in
the Fermi liquid.

It should now be clear that toward the QCP, the effective
Debye temperature $\Omega_t=\omega(k_{\rm max})=
c_Rk_{\rm max}$ {\it goes down to zero}, independently of the
value of the wave number $k_{\rm max}$ characterizing the
cutoff of the TZSM spectrum.  Thus, the necessary condition
$\Omega_t<T$ for emergence of a regime of classical behavior
is always met.  However, another condition must also be satisfied
if there is to exist a well-pronounced classical domain at
extremely low temperature.  Consider that the boson contribution
\beq
F_B=T\int \ln \left( 1-e^{-{ck/ T}}\right)\theta(k-k_{\rm max})
d\upsilon
\label{textf}
\eeq
to the free energy is proportional to some power of $k_{\rm max}$,
depending on the dimensionality of the problem.  The same is true
of the corresponding contributions to kinetic phenomena.
Therefore the extra condition needed is that $k_{\rm max}$ must not to
be too small.  We identify $k_{\rm max}$ with the new
characteristic momentum arising beyond the point of the TPT, namely
the distance $d=p_2-p_1$ between the new sheets of the Fermi
surface.  Indeed, for momenta $p$ situated at distances from
the Fermi surface significantly in excess of $d$, the
single-particle spectrum $\epsilon(p)$ is no longer flat.
The collective spectrum $\omega(k)$ determined from the corresponding
Landau kinetic equation is no longer soft, and consequently
the imaginary part $c_I$ of $c$ becomes of the same order as
$c_R$, barring propagation of the TZSM.

This scenario is illustrated for the key property of resistivity
in electron systems of solids. The kernel of the electron-TZSM
collision integral underlying the resistivity $\varrho(T)$
then contains terms
$ n_o({\bf p}+{\bf k})(1-n_o({\bf p})) N({\bf k})-n_o({\bf p})
(1-n_o({\bf p}+{\bf k}))(1+  N({\bf k}))$ and
$n_o({\bf p}+{\bf k})(1-n_o({\bf p}))(1+ N({\bf k}))- n_o({\bf p})
(1-n_o({\bf p}+{\bf k})) N({\bf k})$,
in which $N({\bf k})$ denotes a nonequilibrium TZSM momentum
distribution and $n_o({\bf p})$, a nonequilibrium electron
momentum distribution of the band that can absorb and emit
the TZSM.  The explicit linearized electron-phonon-like form
of the corresponding component of the collision integral
is\cite{kin,kittel}
\beq
I_{e,ph}\propto \int w({\bf p},{\bf k}) \omega(k)
{\partial N_0(\omega)\over\partial\omega}(\delta n_i-\delta n_f))
\delta(\epsilon_i-\epsilon_f))d\upsilon.
\eeq
In this expression, $w$ is the collision probability,
$N_0(\omega)=[\exp(\omega({\bf k})/T)-1]^{-1}$ is the equilibrium
TZSM momentum distribution, and $\delta n_{i,f}$ stands for the
deviation of the real momentum distribution of the electron band
labeled $o$ from its nonequilibrium counterpart, with
$n_i=n({\bf p})$ and $n_f=n({\bf p}+{\bf k})$.  In the classical
case of the electron-phonon interaction, at $T_D<T$ one has
$ \partial N_0(\omega)/\partial\omega\propto -T/\omega^2$
while all the other factors are $T$-independent,
resulting\cite{kin,kittel} in linear variation of the
resistivity $\varrho(T)$ with $T$.  Based on the analogy
we have established between the roles of phonons and the TZSM,
the resistivity of the strongly correlated electron system must
obey the FL law $\varrho(T)\propto T^2$ only at $T < \Omega_t$.
In the opposite case $\Omega_t<T$, the resistivity exhibits a
{\it linear} dependence on $T$.
Imposition of a magnetic field cannot kill the soft mode of
transverse zero sound as long as the flattening of the
single-particle spectrum responsible for strong depression of
the effective Debye temperature $\Omega_t$ persists.

These results and conclusions are in agreement with experimental
data\cite{stegcol} on the low-temperature resistivity of the doped
heavy-fermion metal YbRh$_2$(Si$_{0.95}$Ge$_{0.05}$)$_2$, data which
indicates that the linear-in-$T$ dependence of the resistivity
is robust down to temperatures as low as 20 mK.  Remarkably,
 the linearity of $\varrho(T)$ continues to hold in external magnetic
fields up to $B\simeq 2T$, far in excess of the critical value
$B_c\simeq 0.3T$ at which this compound undergoes some phase
transition with a hidden order parameter.\cite{stegcol}
A linear $T$ dependence of $\varrho(T)$ is present as well in
the Hertz-Millis spin-density-wave (SDW) scenario for the QCP in 2D
electron systems.\cite{hertz,millis}  However, critical spin
 fluctuations die out at $B>B_c$, since the SDW transition is
suppressed.  Thus the observed behavior of $\varrho(T)$ contradicts
the SDW scenario. Our scenario is in fact compatible with this behavior, since the TZSM spectrum is less sensitive than the structure
of critical spin fluctuations to the magnitude of the magnetic
field.

A linear $T$ dependence of $\varrho(T)$ can also emerge if
light carriers are scattered by heavy bipolarons.\cite{alex,alex2}
However, there is no evidence for the presence of these
quasiparticles in heavy-fermion metals.

Let us now turn to the analysis of the soft TZSM contribution to
the thermopower.  One may recall that in the classical situation,
the phonon-drag thermopower $S_d(T)$ associated with
nonequilibrium phonons is known\cite{barnard,kin,kittel}
to account for a substantial part of the Seebeck coefficient $S(T)$.
The same is true for the class of quantum-critical systems
considered here, except that the domain in which the drag term
$S_d(T)$ contributes appreciably extends down to extremely low
temperatures. Significantly, at $T\to 0$ the drag contribution
increases as $T^3$, whereas at $T>\Omega_t$ it falls off as
$T^{-1}$, producing a bell-like shape\cite{kin,kittel}
of $S_d(T)$ with a sharp maximum at $T\simeq \Omega_t$.
Since the remaining contributions to $S$ are rather smooth, this
salient feature of $S_d$ appears to be responsible for the
change of sign of the full Seebeck coefficient $S(T)$ at extremely
low $T$ observed experimentally\cite{stegtherm} in the heavy-fermion
metal YbRh$_2$Si$_2$, as well as the nonregular behavior of $S(T\to 0)$ found in several heavy-fermion compounds.\cite{behnia}
The TZSM scenario proposed here predicts that the Seebeck coefficient
in YbRh$_2$(Si$_{0.95}$Ge$_{0.05}$)$_2$ will exhibit the same anomalous behavior at magnetic fields substantially exceeding a
corresponding critical value $B_c$.

In summary, we have investigated the conditions promoting the
formation of soft damped collective modes that play the
same role in kinetic phenomena as phonons.  We have shown that
such a damped soft branch belonging to the transverse zero-sound
mode emerges when several bands cross the Fermi surface
simultaneously, with one of the bands subject to a divergence
of the effective mass of carriers. We have discussed prerequisites
for lowering the corresponding Debye temperature $\Omega_t$ and
demonstrated that the inequality $\Omega_t<T$ is met near the quantum
critical point, where a classical regime sets in at extremely low
temperatures.

We thank A. Alexandrov and V. Shaginyan for  stimulating discussions.
This research was supported by the McDonnell Center for the Space
Sciences, by Grants No.~2.1.1/4540 and NS-7235.2010.2
from the Russian Ministry of Education and Science, and by
Grants No.~09-02-01284 and 09-02-00056 from the Russian
Foundation for Basic Research.


\begin{thebibliography}{99}

\bibitem{loh} H.\ v.\ L\"ohneysen, A.\ Rosch, M.\ Vojta, P.\ W\"olfle, Rev.\ Mod.\ Phys.\ {\bf 79}, 1015 (2007).

\bibitem{steglich}P.\ Gegenwart, Q.\ Si, F.\ Steglich, Nature Phys.\ {\bf 4}, 186 (2008).

\bibitem{saunders1} A.~Casey, H.~Patel, J.~Nyeki, B.~P.~Cowan, J.~Saunders, Phys.~Rev.~Lett.\ {\bf 90}, 115301 (2003).

\bibitem{saunders2} M.~Neumann, J.~Nyeki, B.~P.~Cowan, J.~Saunders, Science {\bf 317}, 1356 (2007).

\bibitem{stegacta} P.~Gegenwart, J.~Custers, T.~Tayama, K.~Tenya, C.~Geibel, O.~Trovarelli, F.~Steglich, K.~Neumaier, Acta Phys.\ Pol.\ {\bf B 34}, 323 (2003).

\bibitem{stegcol} J.\ Custers, P.\ Gegenwart, S.\ Geibel,  F.\ Steglich, P.\ Coleman, S.\ Paschen, Phys.\ Rev.\ Lett.\ {\bf 104}, 186402 (2010).

\bibitem{taillefer} L.\ Taillefer, arXiv:1003.2972.

\bibitem{kcsz} V.\ A. Khodel, J.\ W.\ Clark, V.\ R.\ Shaginyan, M.\ V.\ Zverev, JETP Lett.\ {\bf 92}, 585 (2010).

\bibitem{pines} D.~Pines, P.~Nozi\`eres, {\it Theory of Quantum Liquids}, v.~1 (W. A. Benjamin, New York-Amsterdam, 1966).

\bibitem{halat} I.\ M.\ Halatnikov, {\it An Introduction to the Theory of Superfluidity} (Benjamin, New York 1965).

\bibitem{chubukov} D.~ M.~Maslov, A.~V.~Chubukov, Phys.\ Rev.\ B {\bf 81}, 045110 (2010).

\bibitem{bud'ko} S.~L.~Bud'ko, E.~Morosan, P.~C.~Canfield, Phys.\ Rev.\ B {\bf 69}, 014415 (2004); {\bf 71}, 054408 (2005).

\bibitem{kczjetp09} V.\ A.\ Khodel, J.\ W.\ Clark, M.\ V.\ Zverev, JETP Lett.~{\bf 90}, 639 (2009).

\bibitem{ckzjmpb10} J.\ W.\ Clark, V.\ A.\ Khodel, M.\ V.\ Zverev, Int.\ J.\ Mod.\ Phys. B {\bf 24} (2010), in press.

\bibitem{shagh} V.\ R.\ Shaginyan, JETP Lett.\ {\bf 79}, 286 (2004).

\bibitem{prb2005} J.~W.~Clark, V.~A.~Khodel, M.~V.~Zverev, Phys.\ Rev.\ B {\bf 71}, 012401 (2005).

\bibitem{godfrin1} C.~B\"auerle, Yu.\ M.\ Bunkov, A.\ S.\ Chen, S.\ N.\ Fisher, H.\ Godfrin, J.\ Low Temp.\ {\bf 110}, 333 (1998).

\bibitem{pekar}  L.~D.~Landau, S.\ I.\ Pekar, Zh.\ Eksp.\ Teor.\ Fiz.\ {\bf 18}, 419 (1948).

\bibitem{alex} A.\ S.\ Alexandrov, N.\ Mott, {\it Polarons and Bipolarons} (World Scientific, Singapore, 1996).

\bibitem{alexk} A.\ S.\ Alexandrov, P.\  P.\ Kornilovitch, Phys.~Rev.~Lett.\ {\bf 82}, 807 (1999).

\bibitem{lifshitz} I.\ M.\ Lifshitz, Sov. Phys. JETP {\bf 11}, 1130 (1960).

\bibitem{volrev} G.~E.~Volovik, Springer Lecture Notes in Physics {\bf 718}, 31 (2007) [cond-mat/0601372].

\bibitem{zb}  M.~V.~Zverev, M.~Baldo, JETP {\bf 87}, 1129 (1998); J.\ Phys.: Condens.\ Matter {\bf 11},  2059 (1999).

\bibitem{shagp} S.~A.~Artamonov, V.~R.~Shaginyan, Yu.~G.~Pogorelov, JETP Lett.\ {\bf 68}, 942 (1998).

\bibitem{prb2008} V.\ A.\ Khodel, J.\ W.\ Clark, M.\ V.\ Zverev, Phys.\ Rev.\ B {\bf 78}, 075120 (2008).

\bibitem{kin} E.\ M.\ Lifshitz, L.\ P.\ Pitaevskii, {\it Physical Kinetics} (Pergamon, Oxford, 1981).

\bibitem{kittel} C.\ Kittel, {\it Introduction to Solid State Physics} (John Wiley \& Sons, New York, 1996).

\bibitem{hertz} J.\ A.\ Hertz, Phys.\ Rev.\ B {\bf 14}, 1165 (1976).

\bibitem{millis} A.\ J.\ Millis, Phys.\ Rev.\ B {\bf 48}, 7183 (1993).

\bibitem{alex2} A.\ S.\ Alexandrov, Physica C {\bf 274}, 237 ( 1997).

\bibitem{barnard} R.\ D.\ Barnard, {\it Thermoelectricity in Metals and Alloys} (Taylor and Francis, London, 1972).

\bibitem{stegtherm} S.\ Hartmann, N.\ Oeschler, C.\ Krellner, C.\ Geibel, S.\ Paschen, F.\ Steglich, Phys.~Rev.~Lett.~{\bf 104}, 096401 (2010).

\bibitem{behnia} K.\ Behnia, D.\ Jaccard, J.\ Flouquet, J.\ Phys.: Condens.\ Matter {\bf 16}, 5187 (2004).

\end{thebibliography}
\end{document}